# On the Extrapolation of Generative Adversarial Networks for downscaling precipitation extremes in warmer climates

Neelesh Rampal[1,3], Peter B. Gibson[2], Steven Sherwood[3] and Gab Abramowitz[3]

[1]National Institute of Water and Atmospheric Research, Auckland, New Zealand.

[2]National Institute of Water and Atmospheric Research, Wellington, New Zealand

[3] Climate Change Research Centre & ARC Centre of Excellence for Climate Extremes, University of New South Wales, Sydney, Australia.

Corresponding author: Neelesh Rampal (neelesh.rampal@niwa.co.nz)

**Key Points:**

- Deterministic (regression) downscaling methods underestimate future increases in extreme precipitation, even when trained on future climates

- Generative Adversarial Networks tested here better capture these future increases than deterministic methods, even when trained historically

- Generative Adversarial Networks trained on future climates have better historical and future extrapolation skill vs. historical training

**Abstract**

While deep-learning downscaling algorithms can generate fine-scale climate projections cost-effectively, it is still unclear how well they will extrapolate to unobserved climates. We assess the extrapolation capabilities of a deterministic Convolutional Neural Network baseline and a Generative Adversarial Network (GAN) built with this baseline, trained to predict daily precipitation simulated by a Regional Climate Model (RCM). Both approaches emulate future changes in annual mean precipitation well, even when trained on historical data, though training on a future climate improves performance. For extreme precipitation (99.5th percentile), RCM simulations predict a robust end-of-century increase with future warming (~5.8%/°C on average from five simulations). When trained on a future climate, GANs capture 97% of the warming-driven increase in extreme precipitation compared to 65% in a deterministic baseline. Even GANs trained historically capture 77% of this increase. Overall, GANs offer better generalization for downscaling extremes, which is important in applications relying on historical data.

**Plain Language Summary**

The resolution of climate models (~150km) is too coarse for studying the effects of climate change at regional scales. The resolution can be enhanced or 'downscaled' by a physics-based method known as dynamical downscaling, but it is costly and limits the number of climate models that can be downscaled. Deep learning approaches offer a promising and computationally efficient alternative to dynamical downscaling, but it is unclear whether their downscaling of climate models produces plausible and reliable climate projections. We show that one commonly used deep-learning algorithm underestimates future projections of extreme rainfall. However, we show that another algorithm known as a Generative Adversarial Network is better suited for predicting future changes in extreme rainfall and could be useful in similar applications.

1. **Introduction**

Regional Climate Models (RCMs) simulate the impacts of climate change at fine-scales that are often not resolved at the typical spatial resolution of Global Climate Models (GCMs) (e.g., Feser et al., 2011; Giorgi et al., 1994; Prein et al., 2015). However, the large computational cost of RCMs can limit the number of GCMs that can be downscaled, often resulting in under-sampling of model structural and internal variability uncertainty in climate projections, despite its known importance at regional scales (Deser et al., 2012; Deser & Phillips, 2023; Gibson, Rampal, et al., 2024; Hawkins & Sutton, 2009, 2011).

Empirical downscaling algorithms are often several orders of magnitude more computationally efficient than RCMs, making them valuable tools for studying model structural and internal variability uncertainty at regional scales (Maraun et al., 2015; Rampal et al., 2024; Sun et al., 2024; Wilby & Wigley, 1997). Empirical downscaling algorithms are trained to learn relationships between large-scale predictor fields simulated in reanalysis or a GCM (i.e. prognostic variables) and a high-resolution target variable (i.e. precipitation) (see Rampal et al., 2024 for a review). The target variable can be observational data (e.g., daily precipitation) or a simulated variable from an RCM, where the former is known as observational downscaling, and the latter is RCM emulation (Rampal et al., 2024).

The algorithm must be sophisticated enough to capture the spatiotemporal characteristics of the target variable but not too complex so that it poorly generalizes (overfits) when applied out-of-sample (Baño-Medina et al., 2020; Rampal et al., 2022, 2024; Sun et al., 2024). Traditional machine learning (ML) algorithms (e.g. Multilayer Perceptron) are skilful for some problems and datasets but typically require manual feature engineering to incorporate complex and non-linear spatial relationships into the algorithm (i.e. spatial gradients; Gibson et al., 2021). This is usually a limitation when downscaling precipitation, which is often strongly associated with spatial gradients across different scales and complex multivariate relationships across predictor variables (Rampal et al., 2022). This limitation can be overcome using computer-vision ML algorithms such as Convolutional Neural Networks (CNNs), which can automatically learn spatial relationships at different length scales, such as gradients, through their convolutional layers (LeCun et al., 2015).

Computer-vision downscaling algorithms can be trained deterministically or generatively. Deterministic approaches produce a single prediction, whereas generative methods predict an ensemble of possible outcomes. Deterministic approaches are typically trained using regression-based loss functions like mean squared error (MSE), which minimize pixel-wise distances between the predicted and ground truth targets. Generative loss functions, such as in Generative Adversarial Networks (GANs), focus less on pixel-wise accuracy and promote predictions that better characterize the spatial structure and distribution of the target variable (Harris et al., 2022; Leinonen et al., 2021; Rampal et al., 2024.; Vosper et al., 2023).

Before applying a downscaling algorithm to a large ensemble of GCMs, we need to be confident that it will extrapolate to a wide range of future climates, which may be outside the training distribution (Baño-Medina et al., 2021; Boé et al., 2023; Doury et al., 2022, 2024; Gutiérrez et al., 2019; Hernanz et al., 2022a; Rampal et al., 2024; van der Meer et al., 2023). One challenge for algorithms trained only on observational data is that it is not possible to directly evaluate their skill in future climates. However, it is possible to develop pseudoreality, or "model-as-truth" experiments to indirectly test generalization capabilities (i.e. Bano-Medina et al., 2023; Boé et al., 2023; Chadwick et al., 2011; Charles et al., 1999; Doury et al., 2022; Maraun et al., 2015; Rampal et al., 2024). In these experiments, the observable predictor and target variables are replaced with simulated fields from a RCM or "pseudo-observations". It is possible to directly evaluate the algorithm's predictions against "ground truth" model simulations in historical and future climates.

While some studies have used model-as-truth experiments with traditional ML algorithms for specific variables and regions (i.e. Boé et al., 2023; Chadwick et al., 2011; Charles et al., 1999; Hobeichi et al., 2023; Holden et al., 2015; Nishant et al., 2023), few have explored how effectively computer-vision algorithms extrapolate to warmer climates or whether certain algorithms offer advantages (i.e. deterministic vs generative)(Rampal et al., 2024). Generative approaches have outperformed deterministic algorithms in predicting extreme events, capturing high-resolution spatial structures, and preserving statistical properties in historical climates (Addison et al., 2024; Aich et al., 2024; Annau et al., 2023; Mardani et al., 2023; Miralles et al., 2022; Oyama et al., 2023; Rampal et al., 2024.; Saha & Ravela, 2022; Sha et al., 2024). However, generative approaches have been less thoroughly evaluated in future climate contexts and for indices such as extreme precipitation (i.e., 99.5th percentile). In particular, the climate change signal for extreme precipitation may be challenging to learn as there are a limited number of events, and yet there is strong warming-driven signal (Clausius-Clapeyron, ~7%/°C)(i.e. Bao et al., 2017; Dai et al., 2024; Emori & Brown, 2005; Pall et al., 2006.; Pfahl et al., 2017).

Our study focuses on two important gaps in the literature regarding the extrapolation of empirical downscaling algorithms. First, we examine how well relationships learned from a historical period extrapolate to future unobserved climates. We compare two widely used algorithms, a GAN and a deterministic CNN baseline, that use a similar architecture (i.e. convolutional layers) trained in a model-as-truth framework to downscale daily precipitation over New Zealand. We evaluate their accuracy in capturing climate change signals in mean and extreme precipitation. Second, we explore whether training on future vs. only historical periods combined with different-sized training datasets can improve extrapolation skill.

## 2. Materials and Methods

### 2.1 Training and Evaluation Data

This study follows the methodology outlined in Rampal et al. (2024b) and uses a model-as-truth framework for training and evaluation. The predictor and target variables are from the variable-resolution RCM known as the Conformal Cubic Atmospheric Model (CCAM) (Chapman et al., 2023; Gibson et al., 2023; McGregor & Dix, 2008; Thatcher & McGregor, 2009) over the New Zealand region. A detailed evaluation of CCAM for this region is presented in Gibson et al. (2024) and Campbell et al. (2024).

The target variable (psuedo-observations) is daily accumulated (~12km) precipitation ($pr$) from CCAM over the New Zealand region (165°E-184°W, 33°S-51°S), which is logarithmically normalized ($z == \ln[pr + 0.001]$). We use daily averaged coarse-resolution predictor variables from CCAM, which include zonal wind ($U$), meridional wind ($V$), temperature ($T$) and specific humidity ($Q$) at two pressure levels (500mb and 850mb) in the atmosphere. The predictor variable domain has a larger extent than the target variable (151°E-188°W, 26°S-59°S) to prevent information scarcity at the lateral boundaries of the target domain (Bailie et al., 2024; Rampal et al., 2024b., 2022). The predictor variables are re-gridded from 12km to a resolution of 1.5° (~150km) using conservative remapping to be consistent with typical GCM resolution. They are also normalized relative to the mean and standard deviation computed over the entire training dataset (e.g. Bailie et al., 2024; Rampal et al., 2024b., 2022; Rasp et al., 2020).

All experiments are trained using the predictor and target fields from the CCAM downscaled simulation driven by the ACCESS-CM2 GCM. Instead of using predictor fields directly from the GCM, we coarsen CCAM's high-resolution fields (12 km) to the resolution of a typical CMIP6 GCM (1.5°), known as the "perfect framework" for downscaling (Boé et al., 2023; Doury et al., 2022; Rampal et al., 2024; van der Meer et al., 2023). By training and evaluating in the perfect framework, we only isolate the impact of non-stationary future data on downscaling algorithm performance without the confounding influence of transferability between RCM and GCM relationships.

Evaluation is performed on the historical and SSP370 scenarios of four independent CCAM simulations (i.e., not used in training), where the algorithms are directly applied to the coarsened predictor fields for these simulations. The algorithm's predictions are compared to CCAM's (ground truth) precipitation using the metrics discussed in Section 2.5. These CCAM simulations are driven by four GCMs that span a broad range of future warming rates (Gibson et al., 2024). The simulated mean temperature increases by the following amount for the 2070-2099 climatology relative to 1985-2014 over the New Zealand land region: EC-Earth3 (+2.91°C), NorESM2-MM (+1.78°C), AWI-MR-1 (+2.66°C), and CMRM-CM6 (+3.21°C). The training simulation (ACCESS-CM2) increases by +3.49°C over this period.

As implemented in Rampal et al. (2024b), we use two commonly used empirical downscaling algorithms: A deterministic CNN baseline (regression-based) and a residual GAN (generative). Each algorithm is trained with three variations of the CCAM-driven ACCESS-CM2 simulation (historical-only, future-only, and combined historical and future), producing six experiments (Table 1). The future-only training experiment (SSP370) enables us to isolate the effect of training on future data, independent of training size (i.e., by design, the historical and future data sizes are the same).

| Algorithm | Training Data | Period |
|---|---|---|
| Deterministic Baseline | Historical | 1960-2014 (~21,000 days) |
| Deterministic Baseline | Future (SSP370) | 2044-2099 (~21,000 days) |
| Deterministic Baseline | Historical and Future | 1960-2099 (~51,000 days) |

| | (SSP370) | |
|---|---|---|
| Residual GAN | Historical | 1960-2014 |
| Residual GAN | Future (SSP370) | 2044-2099 |
| Residual GAN | Historical and Future (SSP370) | 1960-2099 |

**Table 1:** The six RCM emulator experiments performed in this study.

## 2.2 Deterministic Baseline

The deterministic baseline CNN algorithm is based on the U-Net architecture. The architecture employed is identical to that described in Rampal et al. (2024b), featuring both contracting and expansive pathways, where some intermediate layers are "skip-connected" between the pathways. The U-Net architecture incorporates residual blocks within both contracting and expansive pathways, improving performance and stability compared to traditional convolutional layers (Rampal et al., 2024; Sun et al., 2024). All convolutional layers employ a fixed 3 by 3 kernel size and filter sizes of 32, 64, and 128 within residual blocks of the contracting path (reversed in the expansive path). In total, the U-Net has 1.5 million trainable parameters.

## 2.3 Residual Generative Adversarial Network

GANs involve two models trained to compete: a generator and a critic. The objective of a generator is to fool the critic into thinking the generator's downscaled precipitation (fake simulated data) is, in fact, ground truth precipitation (real data). The critic's objective is to distinguish between real data (i.e. CCAM simulations) and the generator's output (Goodfellow et al., 2014; Mirza & Osindero, 2014). Overall, this competition generally leads to the generator implicitly learning through a powerful loss function that goes beyond traditional pixel-wise comparisons, encouraging the generation of outputs to be distributionally and structurally similar to the real data (i.e. Rampal et al., 2024b., 2024).

There are two main loss functions in training a GAN: the generator loss ($G_{loss}$) and the discriminator (critic) loss. The $G_{loss}$ shown in (1) consists of three components: the Mean Squared Error (MSE), the adversarial loss $G_{Adv}$ and its weight ($\lambda_{adv}$) and an intensity constraint (IC).

$$(1): G_{loss} = MSE(y_{true}, \widehat{y_{pred}}) + \lambda_{adv} * G_{adv}(y_{pred}) + IC(y_{true}, y_{pred})$$

where $IC(y_{true}, y_{pred}) = MSE(Y_{true}^{max}, Y_{pred}^{max})$, and

$$\widehat{y_{pred}} = \frac{1}{8}\sum_i y_{pred_i}.$$

The MSE is calculated using the 8-member ensemble average, as in previous studies (Harris et al., 2022; Rampal et al., 2024b.).

Rampal et al. (2024b) highlighted the importance of modifying the $G_{loss}$ to incorporate an intensity constraint, which produced more accurate precipitation statistics and an improved ensemble spread in the context of weather generation. They also found that when the adversarial loss weight ($\lambda_{adv}$) was between 0.005 and 0.1, the GAN most skilful in the historical period of simulation. In this study, we use a value of $\lambda_{adv} = 0.01$, but we have repeated our experiments with $\lambda_{adv} = 0.005$ and $\lambda_{adv} = 0.1$, with similar results (not shown).

As implemented in Rampal et al. (2024b) and Mardani et al. (2023), instead of predicting precipitation directly, the GAN is trained to predict residuals ($r = \widehat{y_{\lambda_{adv}=0}} - y_{true}$) between a deterministic CNN ($\lambda_{adv} = 0$) and CCAM. The deterministic baseline learns the expectation of all outcomes (predictable component) from CCAM, which are typically smooth in space and time. Thus, the GAN generates plausible fluctuations around this expectation and captures high-frequency variations.

For inference, predictions from the residual GAN are added to the deterministic baseline and inverse transformed ($pr = exp(Y_{\lambda_{adv}=0..0} + Y_{\lambda_{adv}})$-0.001) to produce daily precipitation fields. Both the GAN and the deterministic baseline are applied to five GCMs (the GCM it was trained on and four unobserved GCMs) for the historical period (1985-2014) and the end-of-century period (2070-2099) for the SSP3-70 scenario. For the residual GAN algorithm, we create a 10-member ensemble for each GCM. Each experiment was repeated three times with a different random seed to ensure the consistency of results. Generating a single simulation (one member) of a 30-year daily precipitation (~11,000-time steps) record takes approximately 30 seconds on a single A100 GPU.

2.4 Hyperparameter Tuning

As implemented in previous studies (Gulrajani et al., 2017; Leinonen et al., 2021; Rampal et al., 2024b), the generator and discriminator are trained with an initial learning rate of $2 \times 10^{-4}$, and a batch size of 32. The regression baseline is also trained with a learning rate of $7 \times 10^{-4}$. During training, we decay the learning rate exponentially with decay rates of 0.995 and 0.995 per 1000 iterations, respectively. Each model was trained for 250 epochs, which equates to approximately 14 hours of training on two NVIDIA A100 GPUs with 80GB RAM for the historical and SSP370-only algorithms and 35 hours when training on both the historical and SSP370 run.

2.5 Evaluation Metrics

We assess the historical and future out-of-sample performance using an evaluation framework similar to that proposed by Rampal et al. (2024b). The historical evaluation metrics are the Mean Absolute Error (MAE) in the emulated annual and extreme precipitation (99.5th percentile) climatology relative to CCAM (ground truth) simulations. The historical annual precipitation climatology is the average precipitation rate (mm/day) over 30 years (1986-2014), computed for each grid point individually. The historical climatology of extreme precipitation is also calculated over the same 30-year period but for days with precipitation greater than 1 mm/day. The MAE is only computed for "land" grid points.

It is important to assess the algorithm's skill in preserving the RCMs' future climate change signal, as this can provide insight into whether the algorithm's learnt relationships are physically plausible (see Rampal et al., 2024 for a recent review). We evaluate the emulator's climate change signal for annual and extreme precipitation relative to ground truth CCAM. The climate change signal is defined as the relative change (%) of the future climatology (2070-2099) from SSP370 to the historical climatology (1986-2014). The signal is computed for all land grid points. The climatologies are calculated individually across all 10 ensemble members before computing the climate change signal. Because projections of extreme precipitation are often more robust when regionally averaged (i.e. Pall et al., 2006.), we compute its 1) land-averaged climate change signal (across all land grid points) and 2) error relative to the ground truth by first averaging the signal to a ~200 km resolution (coarsening it by a factor of 16). Similar results were obtained without coarsening (Supplementary Figure S10). The climate change signal error is defined as the MAE in the emulated climate change signal relative to CCAM's. This evaluation is performed for all 5 downscaled GCMs.

## 3. Results

We begin with an overview of the historical and climate change signal skill for annual and extreme precipitation (3.1). Then, we examine the climate change signal at different percentiles (3.2).

3.1 Annual and Extreme Precipitation Performance

For the annual precipitation (Figure 1a) and extreme precipitation (Figure 1b), training on future periods (future-only; SSP370, combined historical and future; hist + SSP370) leads to a lower climate change signal error (MAE) than training on the historical period alone for both algorithms (GAN and deterministic baseline). For annual precipitation, both algorithms appear to be equally skillful in capturing the climate change signal across all training dataset configurations (i.e., historical GAN performs similarly to the historical deterministic baseline). Although training on the future period results to a lower error, all algorithms effectively capture the spatial pattern and magnitude of the climate change signal. This is shown in Figure 2a-f for the multi-model mean (averaged across all GCMs), as all algorithms can capture the drying signal across most of the country well. This result is also consistent across individual GCMs (Supplementary Figures S3-S7).

For extreme precipitation (99.5$^{th}$ percentile), GANs have a smaller climate change signal error than deterministic baseline algorithms (Figure 1b) which is comparable to the improvement of training on the future period. Even historically trained GANs outperform the deterministic baseline that is trained on future periods. Figure 2(l-n) shows that the deterministic baseline algorithms capture some aspects of the spatial patterns of the climate change signal but underestimate the magnitude of changes – which are better captured by GANs (Figure 2h-j). This result is also consistent across individual GCMs (Supplementary Figures S3-S7).

GANs are also more skillful in capturing the historical climatology of annual and extreme precipitation compared to the deterministic baseline, with around half the error for extreme precipitation and around a third for annual precipitation for all training configurations (Figure 1a-b). Interestingly, for both mean and extreme precipitation, GANs trained on both future periods have lower error when assessed on the historical period relative to the same algorithm trained only on historical data. This result holds regardless of training dataset size (SSP370 and historical periods have the same length) between the periods.

We find that training on the future-only period has a similar historical and future performance compared to training with the combined historical and future periods. This implies that there is little benefit from training on the historical period, even though the future period begins only in 2044. This suggests that having a strong trend in the training data may be more important for predicting warming-driven signals in annual and extreme precipitation than having the full range of climates (e.g. combined historical and future).

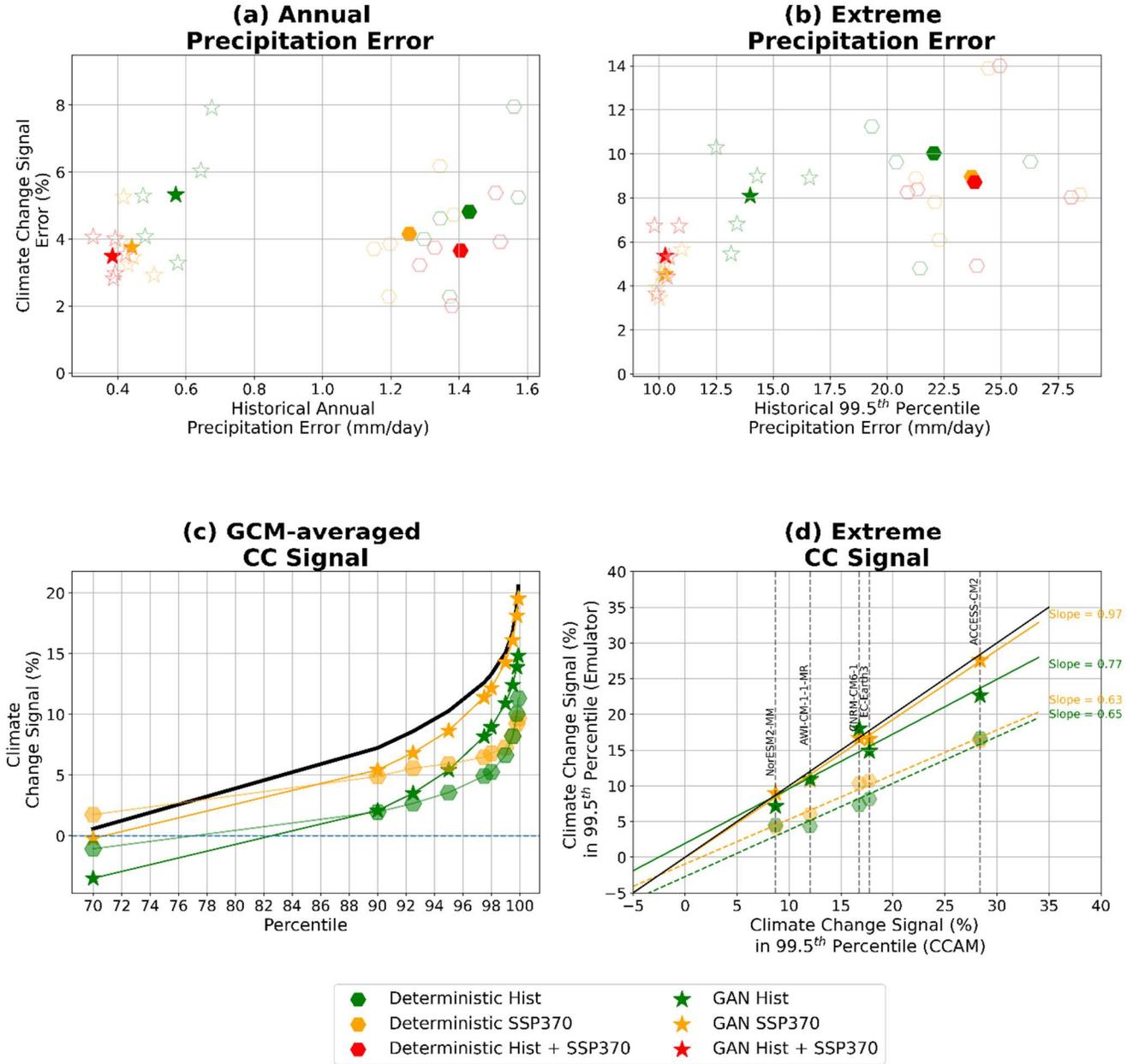

**Figure 1:** Climate change signal MAE (%) versus historical precipitation MAE (mm/day) for (a) mean precipitation and (b) 99.5th percentile of precipitation, where errors are relative to the CCAM ground truth. Star markers represent the GAN algorithms, while circles denote the deterministic baseline. The colours show the data used for training: green for historical, orange for future, and red for historical and future. The filled markers show the average skill across all GCMs, and the unfilled markers show the performance of different GCMs. Panel (c) shows the land-average climate change signal (%) for select percentile values (70th- 99.9th) averaged across five GCMs with CCAM in black. Panel (d) compares CCAM's land-averaged climate change signal at the 99.5th percentile (x-axis) relative to the emulated signal (y-axis) for each GCM.

3.2 Extreme Precipitation Climate Change Signal

The multi-model mean (MME) climate change signal from CCAM varies significantly at different precipitation percentiles. As expected, at the 50th percentile the signal closely matches the annual mean (Section 3.1). This shows a mix of drying and wetting across the country (Supplementary Figure S2), leading to a near-zero or negative signal (Figure 1c). At the 90th percentile we begin to see a weak wetting signal emerge across the country (Supplementary Figure S1). The historically trained GAN and deterministic baseline algorithms show a slight dry bias compared to CCAM but capture the spatial pattern of the wetting signal at the 90th percentile reasonably well (Figure 1c, Supplementary Figure S1). In comparison, the GAN and the deterministic baseline trained on future periods capture both the magnitude and spatial patterns of the signal more accurately. Note that Figures 1c-d show only the land-averaged signal for the future-only GAN, which yields results similar to the GAN trained on the combined historical and future periods (Supplementary Figure S9).

Across the 90th to the 99.9th percentiles, the magnitude of the CCAM wetting signal strongly increases across the region (Figure 1c, Figure 2n). This trend is seen within each driving GCM (Supplementary Figure S3-S8). This shift from a weak or near zero signal at the 90th percentile to a strong wetting signal beyond the 99th percentile has been observed in many tropical and mid-latitude regions (Bai et al., 2024; Bao et al., 2017; Pall et al., 2006.). It is well known that projections of mean and extreme precipitation differ (e.g. Emori and Brown., 2005), where annual precipitation projections are often less spatially homogeneous and do not scale with Clausius Clapeyron (~7% increase per °C). This is thought to be due to annual precipitation being more affected by dynamics (e.g. Emori & Brown, 2005; Gibson, Rampal, et al., 2024) and being constrained by the surface energy budget (Allen & Ingram, 2002; Trenberth et al., 2003).

The historically trained GAN and deterministic baseline exhibit a dry bias in their climate change signals and have relatively similar signals up to the 95th percentile of precipitation. Beyond the 95th percentile, the GAN substantially better captures the wetting signal, although both algorithms underestimate its magnitude compared to CCAM. GANs trained on future periods accurately capture the climate change signal across nearly all percentiles (Figure 1c, Supplementary Figure S9), which are underestimated in the deterministic baselines.

The warming-driven increase in the 99.5th percentile of precipitation in CCAM increases on average at approximately 5.8%/°C based on climate change signal in mean temperature (see Section 2.1 for the temperature signals for each GCM), and higher for the 99.9th percentile (7.2%/°C). The GAN trained on future and historical periods captures about 97% of the average future increase in 99.5th percentile of precipitation, as it closely follows the 1:1 line. However, it slightly underestimates the increase for the ACCESS-CM2 GCM (which has the strongest signal). Conversely, the GAN trained on historical simulations captures about 77% of the warming-driven increase in 99.5th percentile precipitation, where it performs well for most GCMs, except for the ACCESS-CM2 GCM, where it notably underestimates the change. The two deterministic baseline algorithms significantly underestimate the magnitude of the climate change signal, capturing only between 63-65% of the future increase in the 99.5th percentile of precipitation, regardless of being trained on the future or historical periods.

## Annual Climate Change Signal

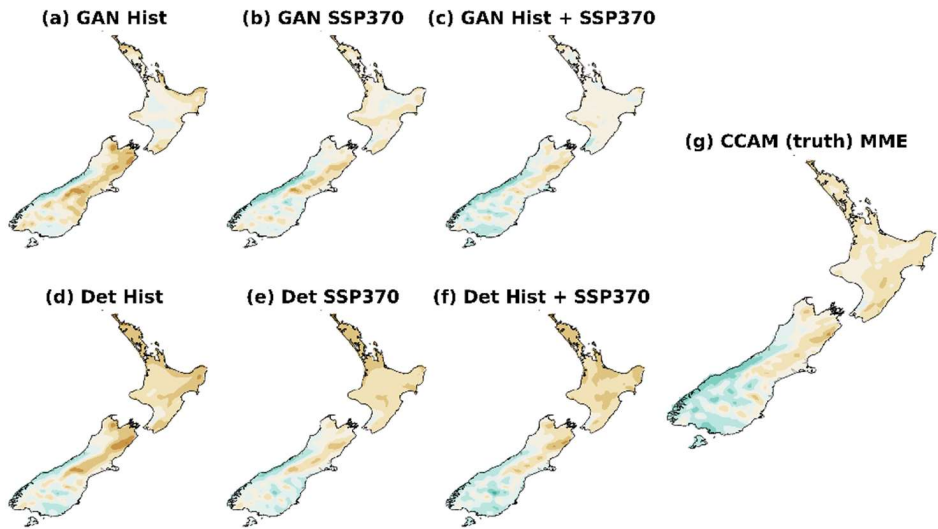

## Extreme Climate Change Signal

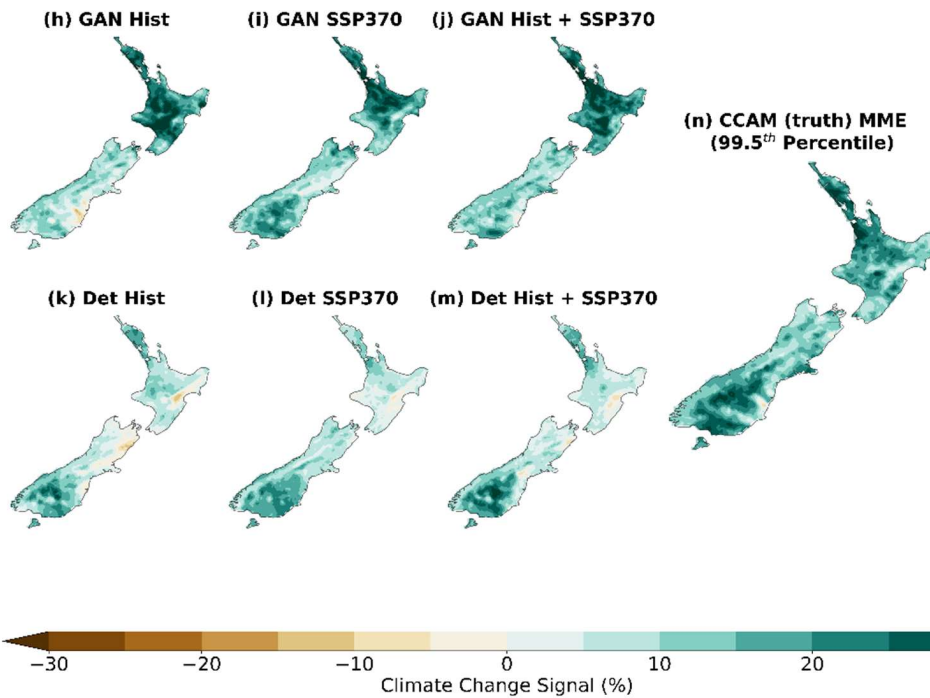

**Figure 2**: The climate change signal (%) of annual precipitation over New Zealand from (a-f) the CNRM-CM6-1 GCM, and (g) CCAM ground truth. (h-n) same but for 99.5th percentile precipitation.

## 4. Discussion

A key finding of our study is that the choice of empirical downscaling algorithm can strongly affect its ability to reproduce CCAM's future climate change signals when trained over the historical period. Both the algorithms capture the annual precipitation signal well, but the deterministic baseline underestimates the increase in extreme precipitation by at least 35%. One should therefore be cautious about the extrapolative capability of deterministic algorithms for similar problems. In contrast, GANs better capture this warming-driven increase in extremes, without any cost to mean performance. Since the architecture is nearly identical between the GAN and the deterministic baseline, the improvements obtained from GANs likely stem from the adversarial loss function. A previous study by Rampal et al., (2024b) found that adversarial loss leads to predictions that accurately capture the spatial structure and distribution of CCAM's precipitation. The additional focus on spatial structure in the adversarial loss function may improve out-of-sample generalization for ill-posed problems like precipitation downscaling. The ill-posed nature of the problem stems from the fact that predictor fields may lack sufficient information for regression-based methods to accurately characterize the target variable, leading to overly smooth predictions (i.e. Bashir et al., 2021; Chen et al., 2022; Cheng et al., 2020; Rampal et al., 2024b., 2024; Wang et al., 2018).

Although GANs capture warming-driven increases in extreme precipitation better than deterministic models when trained on historically, they still capture only about 77% of the total increase from CCAM. This limitation may stem from insufficient diversity in the historical weather states, a challenge noted in previous studies on mean precipitation and temperature (i.e. Boé et al., 2023; Chadwick et al., 2011; Doury et al., 2022). We demonstrated that training on future climates, even with the same length of training record as in the historical period, captures most of the warming-driven increases (97%) in extreme precipitation. This improvement is further supported as GANs trained on future periods have lower historical errors for both mean and extreme precipitation. A similar result was also reported in a related temperature downscaling study (Doury et al., 2022). This suggests that certain future periods may be more valuable for training RCM emulators, and perhaps provides a more diverse range of weather states that allows the algorithm to better extrapolate to unobserved GCMs.

The above result has important implications both for RCM emulation and observational downscaling. First, it suggests that emulators may only require training data from specific time slices in future periods that have a diverse range of weather states, as proposed by Hobeichi et al. (2023). This could reduce the need to generate large datasets of RCM simulations required for training, particularly for costly convection-permitting RCMs. However, further research is required to test this for different problems, regions and for different training simulations. Second, this opens potential hybrid training strategies to enhance extrapolation in algorithms limited to observational data on (i.e. in regions where downscaled simulations are not available to train from). Here, RCM simulations could be used for pre-training and observational data for fine-tuning, as suggested by Rampal et al. (2024). Similar strategies have been implemented or proposed in other applications in climate science (i.e. Beucler et al., 2024; Ham et al., 2019; O'Gorman & Dwyer, 2018; Yuval & O'Gorman, 2020).

## 5. Conclusion

We evaluated the extrapolative capabilities of two empirical downscaling algorithms—a deterministic Convolutional Neural Network baseline and a GAN built on this baseline—in downscaling daily precipitation in a model-as-truth framework. We assessed their ability to extrapolate learned relationships from a historical period to four unobserved future climates, where we measured their ability to predict the climate change signal of annual and extreme precipitation relative to CCAM (truth).

Both algorithms, when trained on historical data alone, appear to capture climate signals well for annual precipitation. However, GANs are significantly more skillful in predicting the signal of extreme precipitation than the deterministic baseline, which significantly underestimates warming-driven increases. Training on the future period improves the GAN's ability to capture climate change signals of mean and extreme precipitation and also improves historical performance. This suggests that a trend in the data may be more valuable for training than a full range of climates (combined historical and future scenario), potentially due to the greater diversity of weather states in future climates.

Overall, GANs capture extreme and annual precipitation changes more effectively, making them superior for downscaling and extrapolating. Further research is required to test whether this result holds true for other architectures and implementations, different regions and different target variables.

*Acknowledgements.*

NR, SS, and GA acknowledge the support of the Australian Research Council Centre of Excellence for Climate Extremes (CLEX; CE170100023). NR and PG received funding from the New Zealand MBIE Endeavour Smart Ideas Fund (C01X2202). The authors would also like to acknowledge the New Zealand eScience Infrastructure for providing access to GPUs. The authors declare no conflicts of interest.

*Open Research*

The code, training and validation datasets are available on Zenodo via Rampal., & Gibson (2024). The code and datasets supporting this study are accessible to the public. The code can also be access via GitHub (https://github.com/nram812/On-the-Extrapolation-of-Generative-Adversarial-Networks-for-downscaling-precipitation-extremes)

## 6. References

Addison, H., Kendon, E., Ravuri, S., Aitchison, L., & Watson, P. A. (2024). *Machine learning emulation of precipitation from km-scale regional climate simulations using a diffusion model* (arXiv:2407.14158). arXiv. https://doi.org/10.48550/arXiv.2407.14158

Aich, M., Hess, P., Pan, B., Bathiany, S., Huang, Y., & Boers, N. (2024). *Conditional diffusion models for downscaling & bias correction of Earth system model precipitation* (arXiv:2404.14416). arXiv. http://arxiv.org/abs/2404.14416

Allen, M. R., & Ingram, W. J. (2002). Constraints on future changes in climate and the hydrologic cycle. *Nature*, *419*(6903), 224–232. https://doi.org/10.1038/nature01092

Annau, N. J., Cannon, A. J., & Monahan, A. H. (2023). Algorithmic Hallucinations of Near-Surface Winds: Statistical Downscaling with Generative Adversarial Networks to Convection-Permitting Scales. *Artificial Intelligence for the Earth Systems*, *2*(4). https://doi.org/10.1175/AIES-D-23-0015.1

Bai, J., Ai, W., Tang, H., & Zhang, Z. (2024). Detectable anthropogenic influence on the changes in structure of precipitation over China using CMIP6 models. *Climate Dynamics*. https://doi.org/10.1007/s00382-024-07368-y

Bailie, T., Koh, Y. S., Rampal, N., & Gibson, P. B. (2024). Quantile-Regression-Ensemble: A Deep Learning Algorithm for Downscaling Extreme Precipitation. *Proceedings of the AAAI Conference on Artificial Intelligence*, *38*(20), Article 20. https://doi.org/10.1609/aaai.v38i20.30193

Bano-Medina, J., Iturbide, M., Fernandez, J., & Gutierrez, J. M. (2023). *Transferability and explainability of deep learning emulators for regional climate model projections: Perspectives for future applications* (arXiv:2311.03378). arXiv. http://arxiv.org/abs/2311.03378

Baño-Medina, J., Manzanas, R., & Gutiérrez, J. M. (2020). Configuration and intercomparison of deep learning neural models for statistical downscaling. *Geoscientific Model Development*, *13*(4), 2109–2124. https://doi.org/10.5194/gmd-13-2109-2020

Baño-Medina, J., Manzanas, R., & Gutiérrez, J. M. (2021). On the suitability of deep convolutional neural networks for continental-wide downscaling of climate change projections. *Climate Dynamics*, *57*(11), 2941–2951. https://doi.org/10.1007/s00382-021-05847-0

Bao, J., Sherwood, S. C., Alexander, L. V., & Evans, J. P. (2017). Future increases in extreme precipitation exceed observed scaling rates. *Nature Climate Change*, *7*(2), 128–132. https://doi.org/10.1038/nclimate3201

Bashir, S. M. A., Wang, Y., Khan, M., & Niu, Y. (2021). A comprehensive review of deep learning-based single image super-resolution. *PeerJ Computer Science*, *7*, e621. https://doi.org/10.7717/peerj-cs.621

Beucler, T., Gentine, P., Yuval, J., Gupta, A., Peng, L., Lin, J., Yu, S., Rasp, S., Ahmed, F., O'Gorman, P. A., Neelin, J. D., Lutsko, N. J., & Pritchard, M. (2024). Climate-invariant machine learning. *Science Advances*, *10*(6), eadj7250. https://doi.org/10.1126/sciadv.adj7250

Boé, J., Mass, A., & Deman, J. (2023). A simple hybrid statistical–dynamical downscaling method for emulating regional climate models over Western Europe. Evaluation, application, and role of added value? *Climate Dynamics*, *61*(1), 271–294. https://doi.org/10.1007/s00382-022-06552-2

Brochet, C., Raynaud, L., Thome, N., Plu, M., & Rambour, C. (2023). Multivariate Emulation of Kilometer-Scale Numerical Weather Predictions with Generative Adversarial Networks: A Proof of Concept. *Artificial Intelligence for the Earth Systems*, *2*(4). https://doi.org/10.1175/AIES-D-23-0006.1

Campbell, I., Gibson, P. B., Stuart, S., Broadbent, A. M., Sood, A., Pirooz, A. A. S., & Rampal, N. (2024). Comparison of three reanalysis-driven regional climate models over New Zealand: Climatology and extreme events. *International Journal of Climatology*, *n/a*(n/a). https://doi.org/10.1002/joc.8578

Chadwick, R., Coppola, E., & Giorgi, F. (2011). An artificial neural network technique for downscaling GCM outputs to RCM spatial scale. *Nonlinear Processes in Geophysics*, *18*(6), 1013–1028. https://doi.org/10.5194/npg-18-1013-2011

Chapman, S., Syktus, J., Trancoso, R., Thatcher, M., Toombs, N., Wong, K. K.-H., & Takbash, A. (2023). Evaluation of Dynamically Downscaled CMIP6-CCAM Models Over Australia. *Earth's Future*, *11*(11), e2023EF003548. https://doi.org/10.1029/2023EF003548

Charles, S., Bates, B., Whetton, P., & Hughes, J. (1999). Validation of downscaling models for changed climate conditions: Case study of southwestern Australia. *Climate Research*, *12*, 1–14. https://doi.org/10.3354/cr012001

Chen, H., He, X., Qing, L., Wu, Y., Ren, C., Sheriff, R. E., & Zhu, C. (2022). Real-world single image super-resolution: A brief review. *Information Fusion*, *79*, 124–145. https://doi.org/10.1016/j.inffus.2021.09.005

Cheng, J., Kuang, Q., Shen, C., Liu, J., Tan, X., & Liu, W. (2020). ResLap: Generating High-Resolution Climate Prediction Through Image Super-Resolution. *IEEE Access*, *8*, 39623–39634. IEEE Access. https://doi.org/10.1109/ACCESS.2020.2974785

Dai, P., Nie, J., Yu, Y., & Wu, R. (2024). Constraints on regional projections of mean and extreme precipitation under warming. *Proceedings of the National Academy of Sciences*, *121*(11), e2312400121. https://doi.org/10.1073/pnas.2312400121

Deser, C., Phillips, A., Bourdette, V., & Teng, H. (2012). Uncertainty in climate change projections: The role of internal variability. *Climate Dynamics*, *38*(3), 527–546. https://doi.org/10.1007/s00382-010-0977-x

Deser, C., & Phillips, A. S. (2023). A range of outcomes: The combined effects of internal variability and anthropogenic forcing on regional climate trends over Europe. *Nonlinear Processes in Geophysics*, *30*(1), 63–84. https://doi.org/10.5194/npg-30-63-2023

Doury, A., Somot, S., & Gadat, S. (2024). On the suitability of a convolutional neural network based RCM-emulator for fine spatio-temporal precipitation. *Climate Dynamics*. https://doi.org/10.1007/s00382-024-07350-8

Doury, A., Somot, S., Gadat, S., Ribes, A., & Corre, L. (2022). Regional climate model emulator based on deep learning: Concept and first evaluation of a novel hybrid downscaling approach. *Climate Dynamics*, *60*(5), 1751–1779. https://doi.org/10.1007/s00382-022-06343-9

Emori, S., & Brown, S. J. (2005). Dynamic and thermodynamic changes in mean and extreme precipitation under changed climate. *Geophysical Research Letters*, *32*(17). https://doi.org/10.1029/2005GL023272

Feser, F., Rockel, B., von Storch, H., Winterfeldt, J., & Zahn, M. (2011). Regional Climate Models Add Value to Global Model Data: A Review and Selected Examples. *Bulletin of the American Meteorological Society*, *92*(9), 1181–1192. https://doi.org/10.1175/2011BAMS3061.1

Gibson, P. B., Chapman, W. E., Altinok, A., Delle Monache, L., DeFlorio, M. J., & Waliser, D. E. (2021). Training machine learning models on climate model output yields skillful interpretable seasonal precipitation forecasts. *Communications Earth & Environment*, *2*(1), 1–13. https://doi.org/10.1038/s43247-021-00225-4

Gibson, P. B., Rampal, N., Dean, S. M., & Morgenstern, O. (2024). Storylines for Future Projections of Precipitation Over New Zealand in CMIP6 Models. *Journal of Geophysical Research: Atmospheres*, *129*(5), e2023JD039664. https://doi.org/10.1029/2023JD039664

Gibson, P. B., Stone, D., Thatcher, M., Broadbent, A., Dean, S., Rosier, S. M., Stuart, S., & Sood, A. (2023). High-Resolution CCAM Simulations Over New Zealand and the South Pacific for the Detection and Attribution of Weather Extremes. *Journal of Geophysical Research: Atmospheres*, *128*(14), e2023JD038530. https://doi.org/10.1029/2023JD038530

Gibson, P. B., Stuart, S., Sood, A., Stone, D., Rampal, N., Lewis, H., Broadbent, A., Thatcher, M., & Morgenstern, O. (2024). Dynamical downscaling CMIP6 models over New

Zealand: Added value of climatology and extremes. *Climate Dynamics*. https://doi.org/10.1007/s00382-024-07337-5

Giorgi, F., Brodeur, C. S., & Bates, G. T. (1994). Regional Climate Change Scenarios over the United States Produced with a Nested Regional Climate Model. *Journal of Climate*, *7*(3), 375–399. https://doi.org/10.1175/1520-0442(1994)007<0375:RCCSOT>2.0.CO;2

Goodfellow, I., Pouget-Abadie, J., Mirza, M., Xu, B., Warde-Farley, D., Ozair, S., Courville, A., & Bengio, Y. (2014). Generative Adversarial Nets. *Advances in Neural Information Processing Systems*, *27*. https://proceedings.neurips.cc/paper_files/paper/2014/hash/5ca3e9b122f61f8f06494c97b 1afccf3-Abstract.html

Gulrajani, I., Ahmed, F., Arjovsky, M., Dumoulin, V., & Courville, A. C. (2017). Improved Training of Wasserstein GANs. *Advances in Neural Information Processing Systems*, *30*. https://proceedings.neurips.cc/paper_files/paper/2017/hash/892c3b1c6dccd52936e27cbd0 ff683d6-Abstract.html

Gutiérrez, J. M., Maraun, D., Widmann, M., Huth, R., Hertig, E., Benestad, R., Roessler, O., Wibig, J., Wilcke, R., Kotlarski, S., San Martín, D., Herrera, S., Bedia, J., Casanueva, A., Manzanas, R., Iturbide, M., Vrac, M., Dubrovsky, M., Ribalaygua, J., … Pagé, C. (2019). An intercomparison of a large ensemble of statistical downscaling methods over Europe: Results from the VALUE perfect predictor cross-validation experiment. *International Journal of Climatology*, *39*(9), 3750–3785. https://doi.org/10.1002/joc.5462

Ham, Y.-G., Kim, J.-H., & Luo, J.-J. (2019). Deep learning for multi-year ENSO forecasts. *Nature*, *573*(7775), Article 7775. https://doi.org/10.1038/s41586-019-1559-7

Harris, L., McRae, A. T. T., Chantry, M., Dueben, P. D., & Palmer, T. N. (2022). A Generative Deep Learning Approach to Stochastic Downscaling of Precipitation Forecasts. *Journal of Advances in Modeling Earth Systems*, *14*(10), e2022MS003120. https://doi.org/10.1029/2022MS003120

Hawkins, E., & Sutton, R. (2009). The Potential to Narrow Uncertainty in Regional Climate Predictions. *Bulletin of the American Meteorological Society*, *90*(8), 1095–1108. https://doi.org/10.1175/2009BAMS2607.1

Hawkins, E., & Sutton, R. (2011). The potential to narrow uncertainty in projections of regional precipitation change. *Climate Dynamics*, *37*(1–2), 407–418. https://doi.org/10.1007/s00382-010-0810-6

Hernanz, A., García-Valero, J. A., Domínguez, M., & Rodríguez-Camino, E. (2022a). A critical view on the suitability of machine learning techniques to downscale climate change projections: Illustration for temperature with a toy experiment. *Atmospheric Science Letters*, *23*(6), e1087. https://doi.org/10.1002/asl.1087

Hernanz, A., García-Valero, J. A., Domínguez, M., & Rodríguez-Camino, E. (2022b). Evaluation of statistical downscaling methods for climate change projections over Spain: Future conditions with pseudo reality (transferability experiment). *International Journal of Climatology*, *42*(7), 3987–4000. https://doi.org/10.1002/joc.7464

Hobeichi, S., Nishant, N., Shao, Y., Abramowitz, G., Pitman, A., Sherwood, S., Bishop, C., & Green, S. (2023). Using Machine Learning to Cut the Cost of Dynamical Downscaling. *Earth's Future*, *11*(3), e2022EF003291. https://doi.org/10.1029/2022EF003291

Holden, P. B., Edwards, N. R., Garthwaite, P. H., & Wilkinson, R. D. (2015). Emulation and interpretation of high-dimensional climate model outputs. *Journal of Applied Statistics*, *42*(9), 2038–2055. https://doi.org/10.1080/02664763.2015.1016412

LeCun, Y., Bengio, Y., & Hinton, G. (2015). Deep learning. *Nature*, *521*(7553), 436–444. https://doi.org/10.1038/nature14539

Leinonen, J., Nerini, D., & Berne, A. (2021). Stochastic Super-Resolution for Downscaling Time-Evolving Atmospheric Fields With a Generative Adversarial Network. *IEEE Transactions on Geoscience and Remote Sensing*, *59*(9), 7211–7223. IEEE Transactions on Geoscience and Remote Sensing. https://doi.org/10.1109/TGRS.2020.3032790

Maraun, D., Widmann, M., Gutiérrez, J. M., Kotlarski, S., Chandler, R. E., Hertig, E., Wibig, J., Huth, R., & Wilcke, R. A. I. (2015). VALUE: A framework to validate downscaling approaches for climate change studies. *Earth's Future*, *3*(1), 1–14. https://doi.org/10.1002/2014EF000259

Mardani, M., Brenowitz, N., Cohen, Y., Pathak, J., Chen, C.-Y., Liu, C.-C., Vahdat, A., Kashinath, K., Kautz, J., & Pritchard, M. (2023). *Generative Residual Diffusion Modeling for Km-scale Atmospheric Downscaling* (arXiv:2309.15214). arXiv. http://arxiv.org/abs/2309.15214

McGregor, J. L., & Dix, M. R. (2008). An Updated Description of the Conformal-Cubic Atmospheric Model. In K. Hamilton & W. Ohfuchi (Eds.), *High Resolution Numerical Modelling of the Atmosphere and Ocean* (pp. 51–75). Springer. https://doi.org/10.1007/978-0-387-49791-4_4

Miralles, O., Steinfeld, D., Martius, O., & Davison, A. C. (2022). Downscaling of Historical Wind Fields over Switzerland Using Generative Adversarial Networks. *Artificial Intelligence for the Earth Systems*, *1*(4). https://doi.org/10.1175/AIES-D-22-0018.1

Mirza, M., & Osindero, S. (2014). *Conditional Generative Adversarial Nets* (arXiv:1411.1784). arXiv. https://doi.org/10.48550/arXiv.1411.1784

Nishant, N., Hobeichi, S., Sherwood, S. C., Abramowitz, G., Shao, Y., Bishop, C., & Pitman, A. J. (2023). Comparison of a novel machine learning approach with dynamical downscaling for Australian precipitation. *Environmental Research Letters*. https://doi.org/10.1088/1748-9326/ace463

O'Gorman, P. A., & Dwyer, J. G. (2018). Using Machine Learning to Parameterize Moist Convection: Potential for Modeling of Climate, Climate Change, and Extreme Events. *Journal of Advances in Modeling Earth Systems*, *10*(10), 2548–2563. https://doi.org/10.1029/2018MS001351

Oyama, N., Ishizaki, N. N., Koide, S., & Yoshida, H. (2023). Deep generative model super-resolves spatially correlated multiregional climate data. *Scientific Reports*, *13*(1), 5992. https://doi.org/10.1038/s41598-023-32947-0

Pall, P., Allen, M. R., & Stone, D. A. (2006.). *Testing the Clausius–Clapeyron constraint on changes in extreme precipitation under CO2 warming*.

Pfahl, S., O'Gorman, P. A., & Fischer, E. M. (2017). Understanding the regional pattern of projected future changes in extreme precipitation. *Nature Climate Change*, *7*(6), Article 6. https://doi.org/10.1038/nclimate3287

Prein, A. F., Langhans, W., Fosser, G., Ferrone, A., Ban, N., Goergen, K., Keller, M., Tölle, M., Gutjahr, O., Feser, F., Brisson, E., Kollet, S., Schmidli, J., van Lipzig, N. P. M., & Leung, R. (2015). A review on regional convection-permitting climate modeling: Demonstrations, prospects, and challenges. *Reviews of Geophysics*, *53*(2), 323–361. https://doi.org/10.1002/2014RG000475

Price, I., & Rasp, S. (2022). Increasing the accuracy and resolution of precipitation forecasts using deep generative models. *Proceedings of The 25th International Conference on Artificial Intelligence and Statistics*, 10555–10571. https://proceedings.mlr.press/v151/price22a.html

Rampal, N., & Gibson, P. (2024). *Dataset for "On the Extrapolation of Generative Adversarial Networks for downscaling precipitation extremes in warmer climates"* [Dataset]. Zenodo. https://doi.org/10.5281/zenodo.13755688

Rampal, N., Gibson, P. B., Sherwood, S., Abramowitz, G., & Hobeichi, S. (2024). *A Reliable Generative Adversarial Network Approach for Climate Downscaling and Weather Generation*. Retrieved August 31, 2024, from https://www.authorea.com/users/449031/articles/848420-a-robust-generative-adversarial-network-approach-for-climate-downscaling-and-weather-generation

Rampal, N., Gibson, P. B., Sood, A., Stuart, S., Fauchereau, N. C., Brandolino, C., Noll, B., & Meyers, T. (2022). High-resolution downscaling with interpretable deep learning: Rainfall extremes over New Zealand. *Weather and Climate Extremes*, *38*, 100525. https://doi.org/10.1016/j.wace.2022.100525

Rampal, N., Hobeichi, S., Gibson, P. B., Baño-Medina, J., Abramowitz, G., Beucler, T., González-Abad, J., Chapman, W., Harder, P., & Gutiérrez, J. M. (2024). Enhancing Regional Climate Downscaling through Advances in Machine Learning. *Artificial Intelligence for the Earth Systems*, *3*(2). https://doi.org/10.1175/AIES-D-23-0066.1

Rasp, S., Dueben, P. D., Scher, S., Weyn, J. A., Mouatadid, S., & Thuerey, N. (2020). WeatherBench: A Benchmark Data Set for Data-Driven Weather Forecasting. *Journal of Advances in Modeling Earth Systems*, *12*(11). https://doi.org/10.1029/2020MS002203

Saha, A., & Ravela, S. (2022). *Downscaling Extreme Rainfall Using Physical-Statistical Generative Adversarial Learning* (arXiv:2212.01446). arXiv. http://arxiv.org/abs/2212.01446

Sha, J., Chen, X., Chang, Y., Zhang, M., & Li, X. (2024). A spatial weather generator based on conditional deep convolution generative adversarial nets (cDCGAN). *Climate Dynamics*, *62*(2), 1275–1290. https://doi.org/10.1007/s00382-023-06971-9

Sun, Y., Deng, K., Ren, K., Liu, J., Deng, C., & Jin, Y. (2024). Deep learning in statistical downscaling for deriving high spatial resolution gridded meteorological data: A systematic review. *ISPRS Journal of Photogrammetry and Remote Sensing*, *208*, 14–38. https://doi.org/10.1016/j.isprsjprs.2023.12.011

Thatcher, M., & McGregor, J. L. (2009). Using a Scale-Selective Filter for Dynamical Downscaling with the Conformal Cubic Atmospheric Model. *Monthly Weather Review*, *137*(6), 1742–1752. https://doi.org/10.1175/2008MWR2599.1

Trenberth, K. E., Dai, A., Rasmussen, R. M., & Parsons, D. B. (2003). *The Changing Character of Precipitation*. https://doi.org/10.1175/BAMS-84-9-1205

van der Meer, M., de Roda Husman, S., & Lhermitte, S. (2023). Deep Learning Regional

    Climate Model Emulators: A Comparison of Two Downscaling Training Frameworks.

    *Journal of Advances in Modeling Earth Systems*, *15*(6), e2022MS003593.

    https://doi.org/10.1029/2022MS003593

Vosper, E., Watson, P., Harris, L., McRae, A., Santos-Rodriguez, R., Aitchison, L., & Mitchell,

    D. (2023). Deep Learning for Downscaling Tropical Cyclone Rainfall to Hazard-

    Relevant Spatial Scales. *Journal of Geophysical Research: Atmospheres*, *128*(10),

    e2022JD038163. https://doi.org/10.1029/2022JD038163

Wang, X., Yu, K., Wu, S., Gu, J., Liu, Y., Dong, C., Loy, C. C., Qiao, Y., & Tang, X. (2018).

    *ESRGAN: Enhanced Super-Resolution Generative Adversarial Networks*

    (arXiv:1809.00219). arXiv. http://arxiv.org/abs/1809.00219

Wilby, R. L., & Wigley, T. M. L. (1997). Downscaling general circulation model output: A

    review of methods and limitations. *Progress in Physical Geography: Earth and

    Environment*, *21*(4), 530–548. https://doi.org/10.1177/030913339702100403

Yuval, J., & O'Gorman, P. A. (2020). Stable machine-learning parameterization of subgrid

    processes for climate modeling at a range of resolutions. *Nature Communications*, *11*(1),

    3295. https://doi.org/10.1038/s41467-020-17142-3

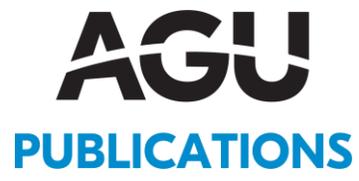

**On the Extrapolation of Generative Adversarial Networks for downscaling precipitation extremes in warmer climates**

Neelesh Rampal[1,3], Peter B. Gibson[2], Steven Sherwood[3] and Gab Abramowitz[3]

[1]National Institute of Water and Atmospheric Research, Auckland, New Zealand.

[2]National Institute of Water and Atmospheric Research, Wellington, New Zealand

[3] Climate Change Research Centre & ARC Centre of Excellence for Climate Extremes, University of New South Wales, Sydney, Australia.

**Contents of this file**

## Annual Climate Change Signal

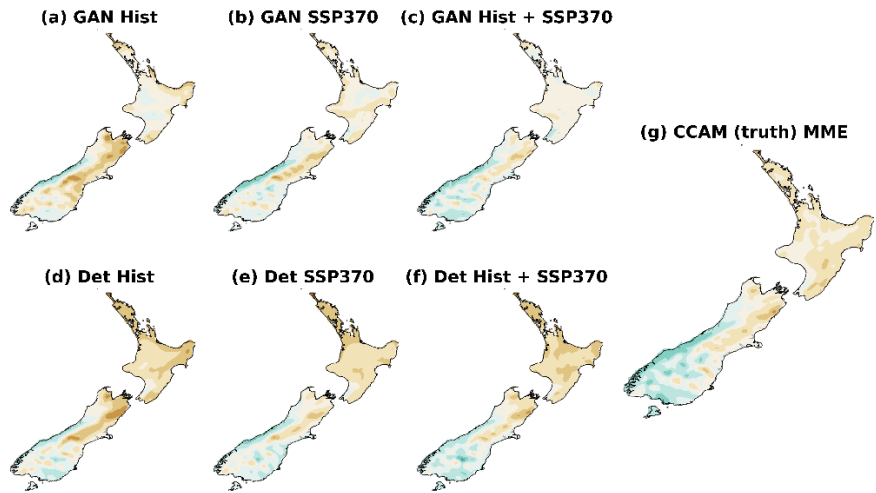

## 90th Percentile Climate Change Signal

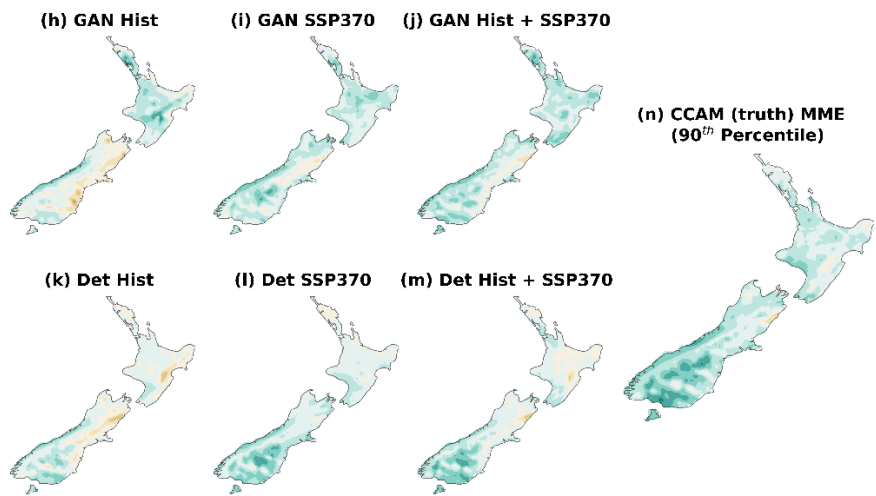

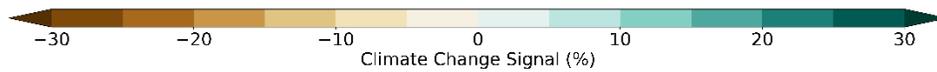

**Figure S1:** Same as Fig. 2 of main text, but (h-n) is computed for the 90th percentile of precipitation.

## Annual Climate Change Signal

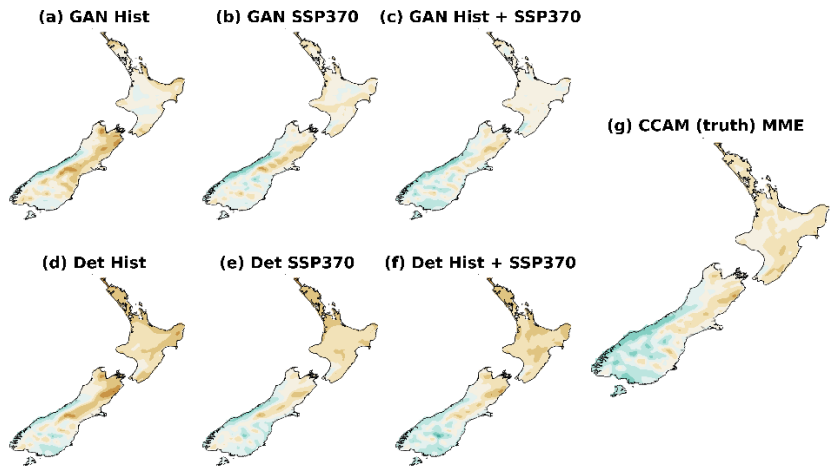

## Median Climate Change Signal

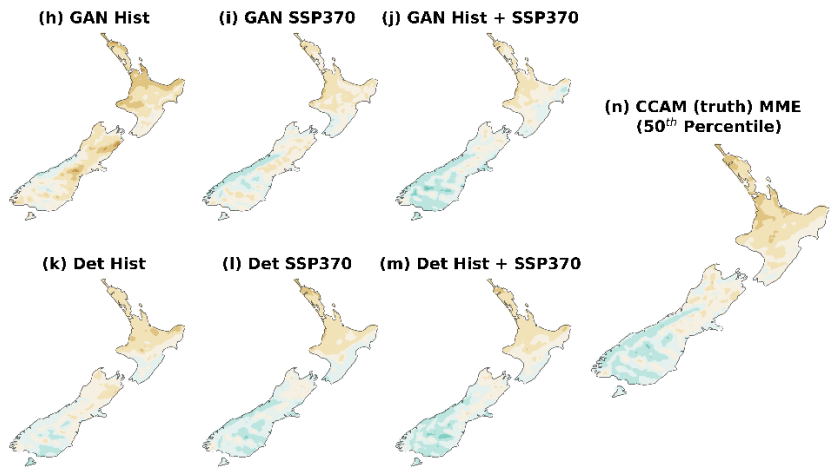

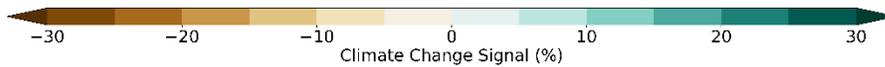

**Figure S2:** Same as Fig. 2 of main text, but (h-n) is computed for the 50th percentile (median) of precipitation.

## Annual Climate Change Signal

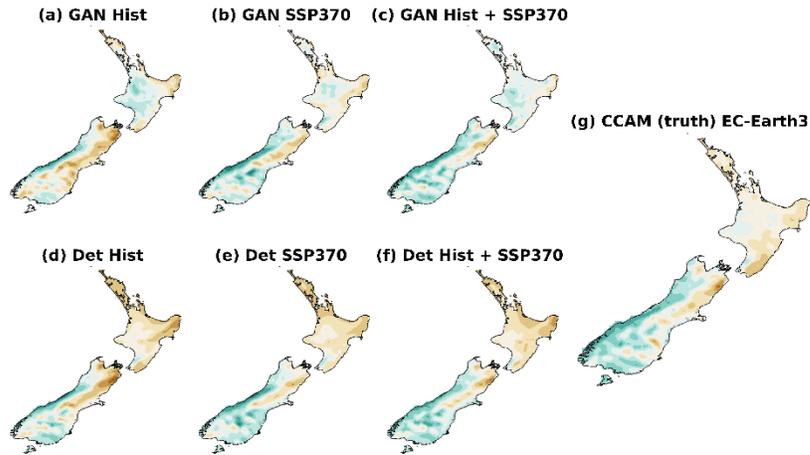

## Extreme Climate Change Signal

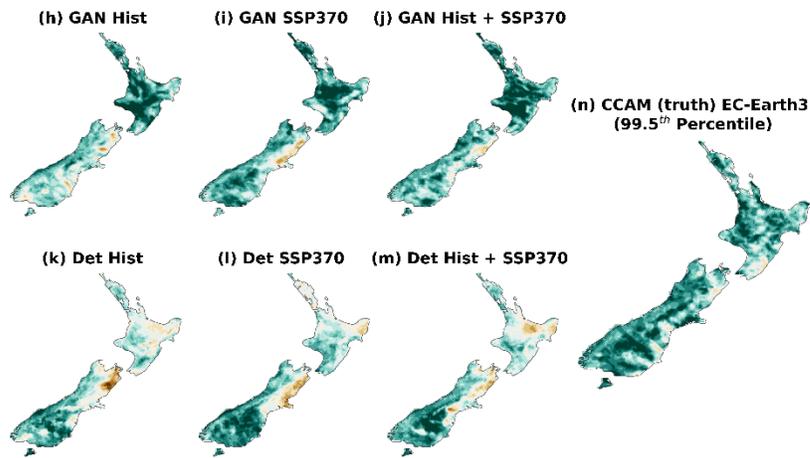

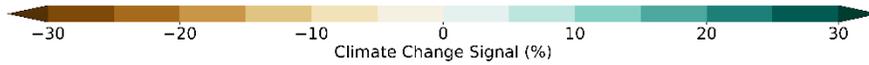

**Figure S3:** As in Fig. 2 of main text, except for the EC-Earth3 driving GCM.

# Annual Climate Change Signal

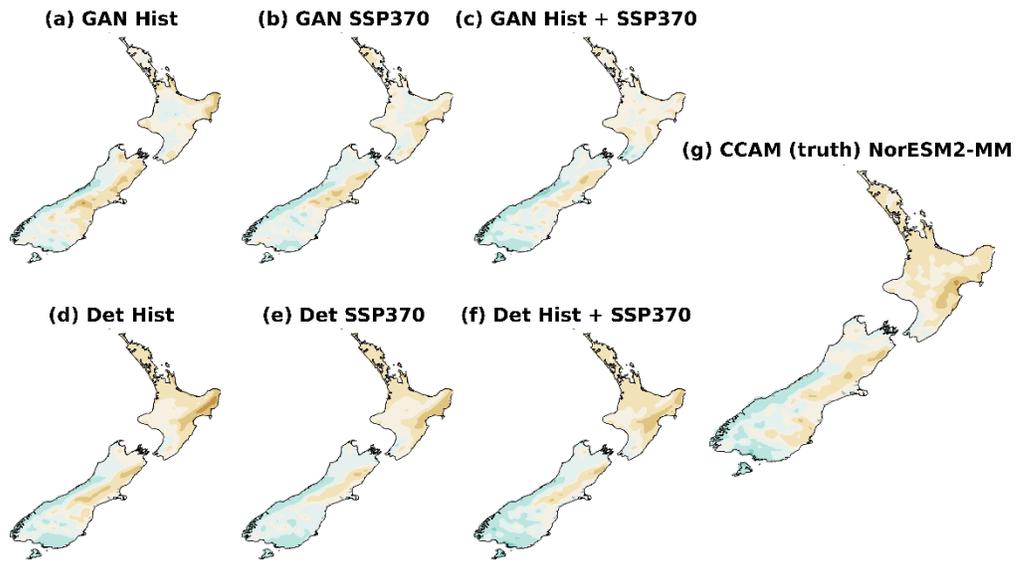

**(a) GAN Hist**   **(b) GAN SSP370**   **(c) GAN Hist + SSP370**

**(g) CCAM (truth) NorESM2-MM**

**(d) Det Hist**   **(e) Det SSP370**   **(f) Det Hist + SSP370**

# Extreme Climate Change Signal

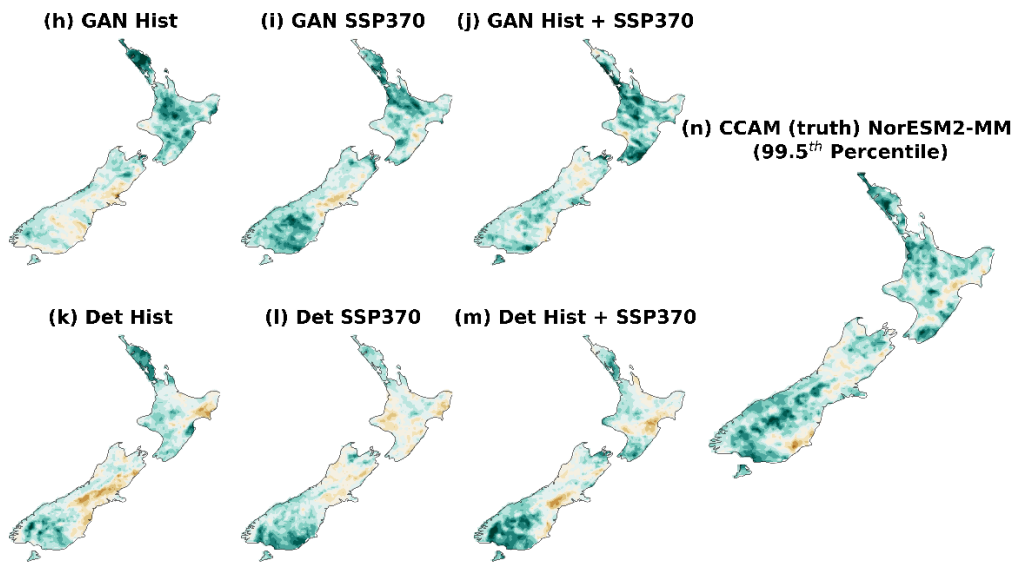

**(h) GAN Hist**   **(i) GAN SSP370**   **(j) GAN Hist + SSP370**

**(n) CCAM (truth) NorESM2-MM**
**(99.5$^{th}$ Percentile)**

**(k) Det Hist**   **(l) Det SSP370**   **(m) Det Hist + SSP370**

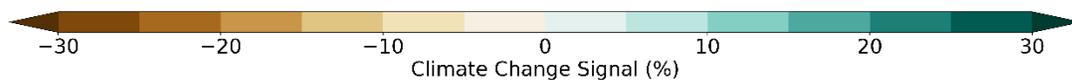

−30   −20   −10   0   10   20   30
Climate Change Signal (%)

**Figure S4:** Same as Figure S3 but for the NorESM2-MM GCM.

# Annual Climate Change Signal

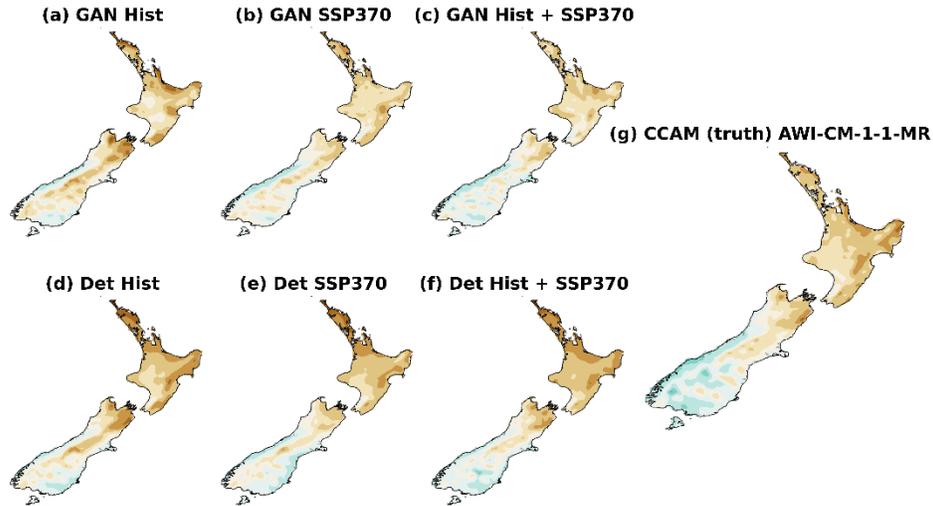

# Extreme Climate Change Signal

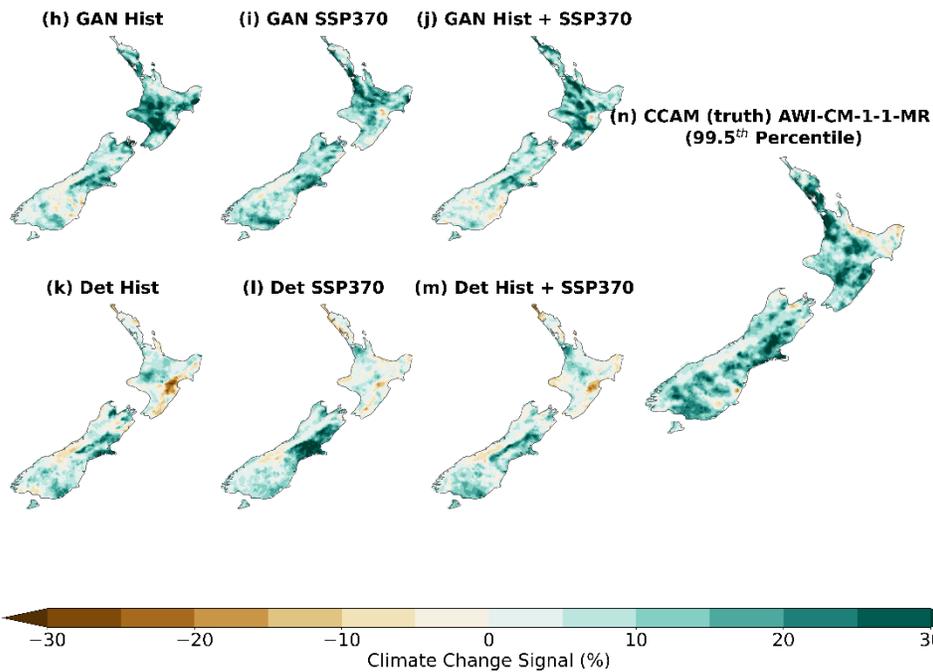

**Figure S5:** (a-g) Same as Figure S3 but for the AWI-CM-1-1MR GCM.

## Annual Climate Change Signal

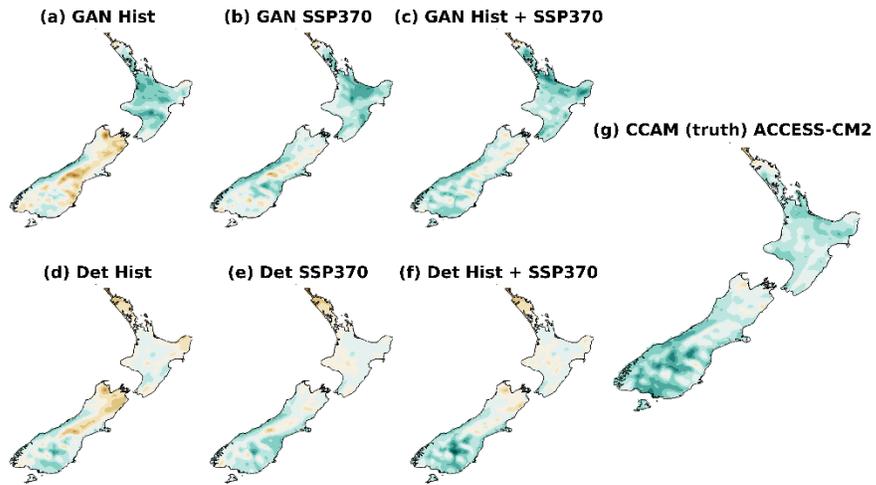

## Extreme Climate Change Signal

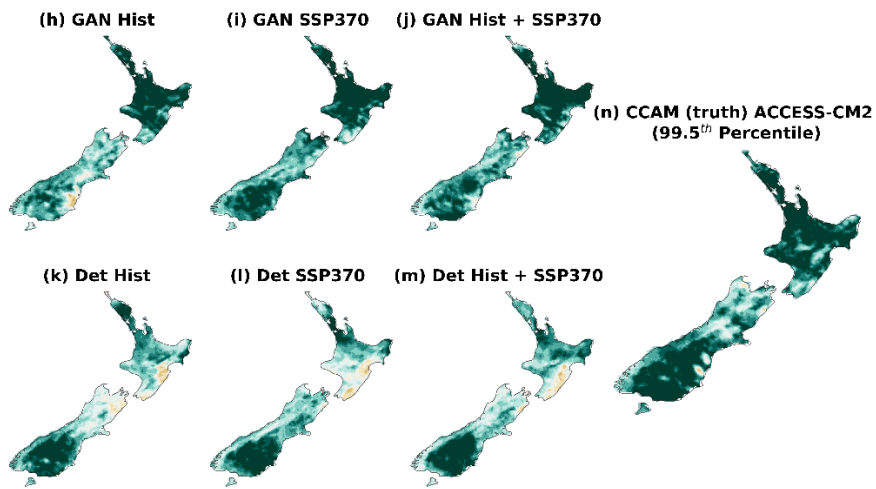

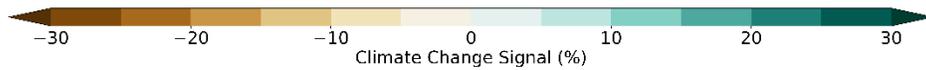

**Figure S6:** (a-g) Same as Figure S3 but for the ACCESS-CM2 GCM.

# Annual Climate Change Signal

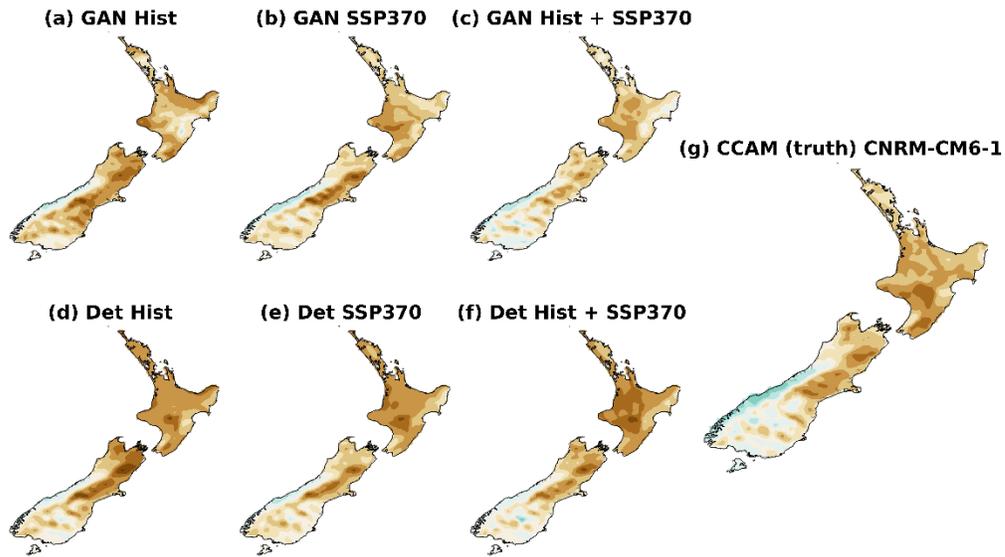

# Extreme Climate Change Signal

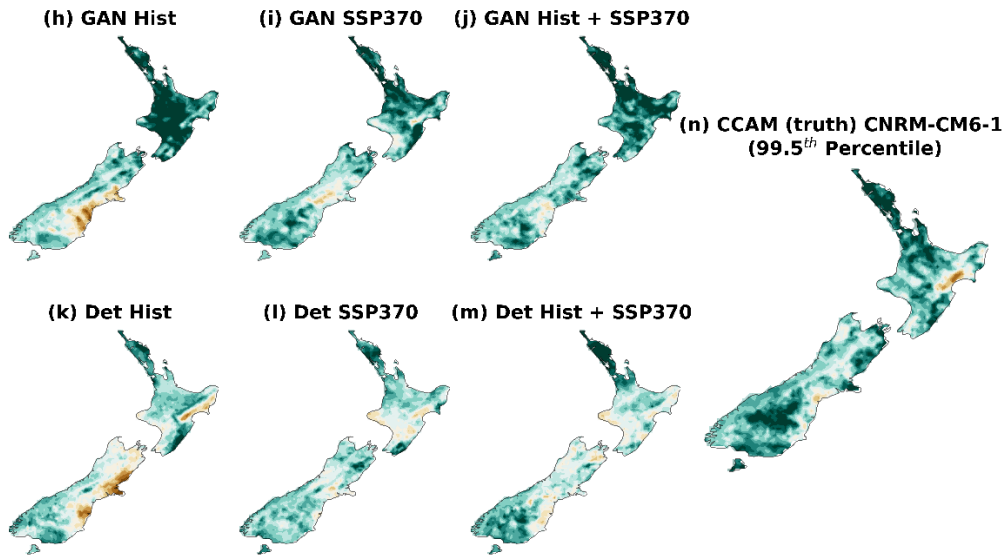

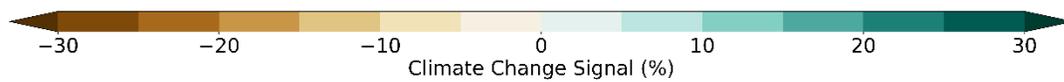

**Figure S7:** (a-g) Same as Figure S3 but for the CNRM-CM6-1 GCM.

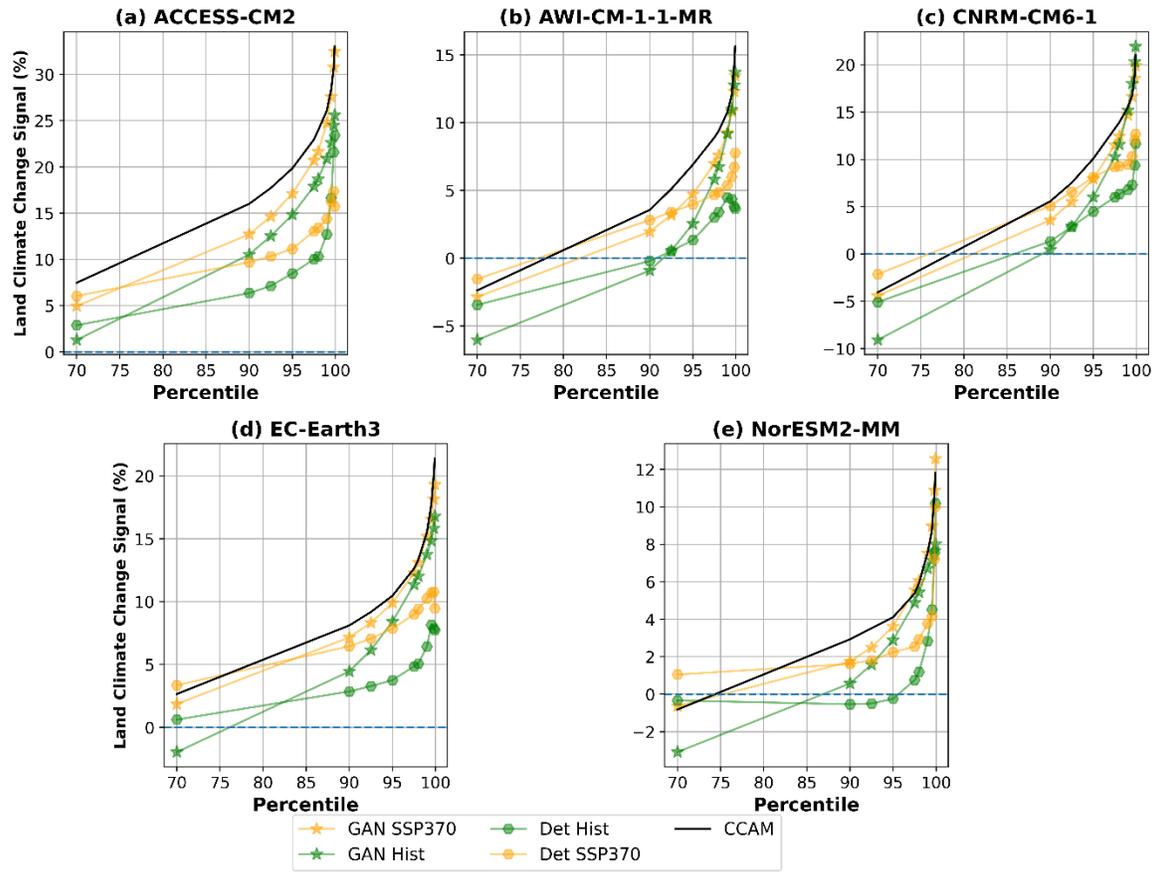

**Figure S8** (a-e) The land-average climate change signal (%) as a function of percentile (90th- 99.9th) averaged for each GCM. The black curve represents the ground truth signal from CCAM.

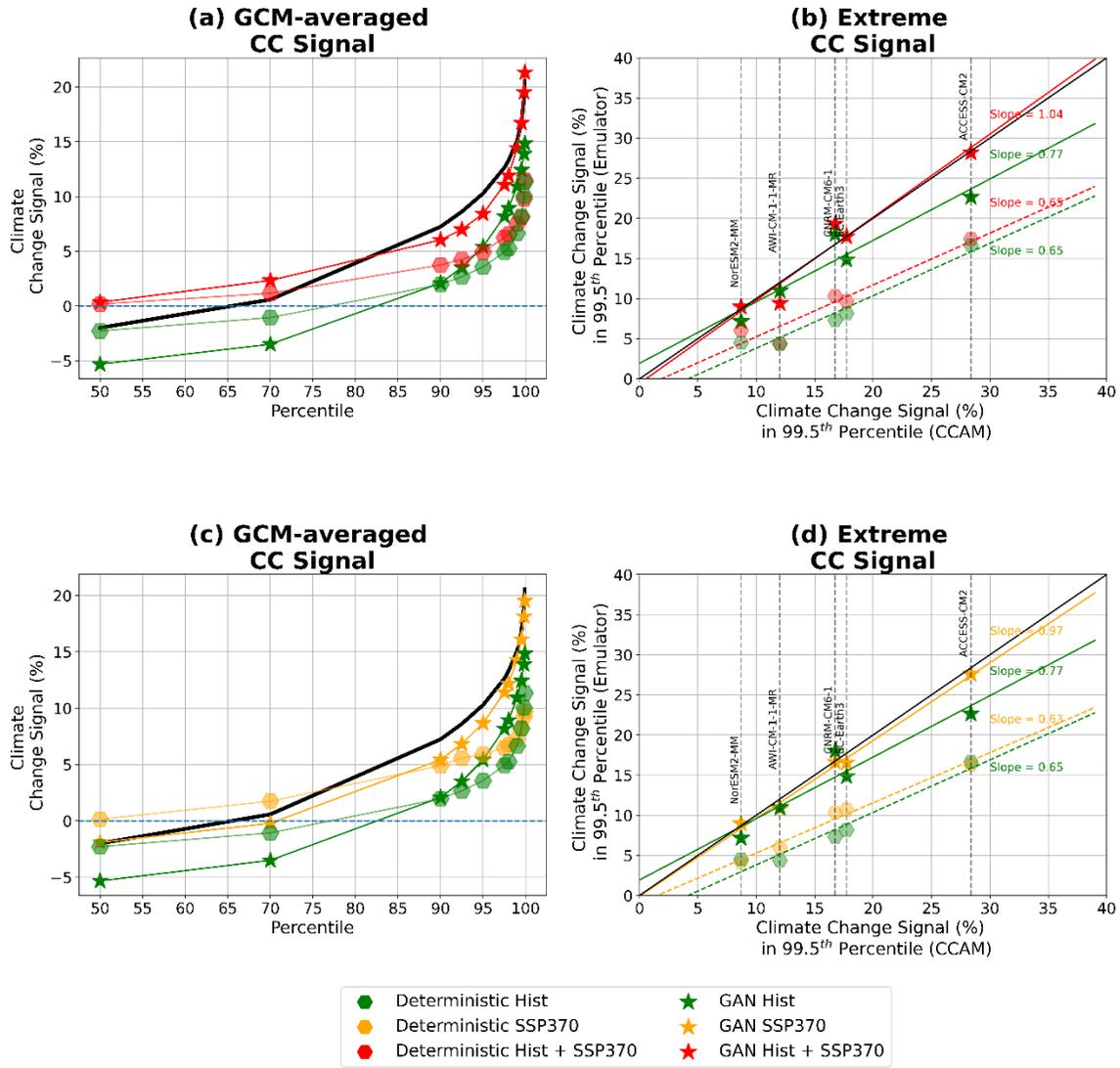

**Figure S9**: Land averaged climate change signal (%) as a function of percentile for the future only (a), and the historical + future (c) GANs and deterministic baseline algorithms. (b-d) The land-averaged climate change signal at the 99.5[th] percentile in ground truth (x-axis) relative to the emulated signal (y-axis).

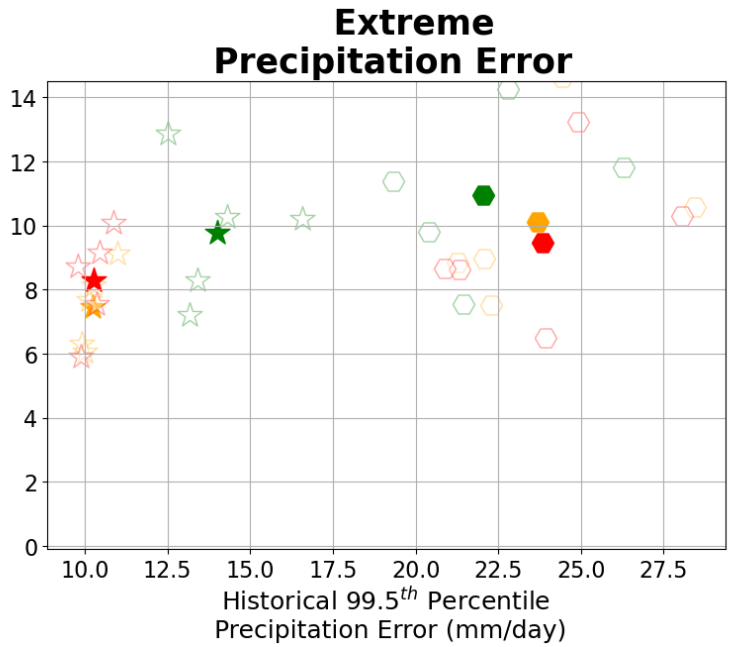

**Figure S10**: Figure 1b, but using pixel-wise MAE as a pose to coarsening the climate change signal to ~200km.